\documentclass{JHEP3}

\pdfoutput=1

\usepackage{graphicx}
\usepackage{amssymb}
\usepackage{slashed}
\usepackage{amsmath}
\usepackage{url}
\usepackage{epsfig}

\def\nnd{\end{document}}

\def\be{\begin{equation}}
\def\ee{\end{equation}}

\newcommand{\bea}{\begin{eqnarray}}
\newcommand{\eea}{\end{eqnarray}}
\newcommand{\bwt}{\begin{widetext}}
\newcommand{\ewt}{\end{widetext}}

\def\u
\def\hZ{\widehat Z}

\def\eed{\end{document}}

\def\m_z{m_{\textrm {Z}}}

\renewcommand{\u}{\rm{u}}

\def\be{\beta}

\def\rm#1{\textrm{#1}}

\def\beq{\begin{equation}}
\def\eeq{\end{equation}}
\def\bea{\begin{eqnarray}}
\def\eea{\end{eqnarray}}

\let\jnfont=\rm
\def\NPB#1,{{\jnfont Nucl.\ Phys.\ B }{\bf #1},}
\def\PLB#1,{{\jnfont Phys.\ Lett.\ B }{\bf #1},}
\def\EPJC#1,{{\jnfont Eur.\ Phys.\ Jour.\ C }{\bf #1},}
\def\PRD#1,{{\jnfont Phys.\ Rev.\ D }{\bf #1},}
\def\PRL#1,{{\jnfont Phys.\ Rev.\ Lett.\ }{\bf #1},}
\def\MPLA#1,{{\jnfont Mod.\ Phys.\ Lett.\ A }{\bf #1},}
\def\JPG#1,{{\jnfont J.\ Phys.\ G }{\bf #1},}
\def\CTP#1,{{\jnfont Commun.\ Theor.\ Phys.\ }{\bf #1},}

\def\NPPS#1,{{\jnfont Nucl.\ Phys.\ Proc.\ Suppl.\ }{\bf #1},}

\DeclareRobustCommand{\Sec}[1]{Sec.~\ref{#1}}

\title{Searching for heavy charged Higgs boson with jet substructure at the LHC}
\author{Shuo Yang$^{a,b}$, Qi-Shu Yan$^c$
\\ $^{a}$ Physics Department, Dalian University, Dalian, 116622, P.R. China  \\ $^{b}$ Center for High-Energy Physics, Peking University, Beijing, 100871, P.R. China
\\$^{c}$ College of Physics Sciences, Graduate University of Chinese Academy of Sciences, Beijing 100039, P.R. China
\\E-mail: \email{yangshuo@dlu.edu.cn, yanqishu@gucas.ac.cn}
 }

\abstract{We study the heavy charged Higgs boson (from 800 GeV to 1500 GeV in this study) in production associated with a top quark at the LHC with the collision energy $\sqrt{s}=14$ TeV. Such a heavy charged Higgs boson can dominantly decay into a top quark and a bottom quark due to its large Yukawa couplings, like in MSSM. To suppress background events and to confirm the signal, we reconstruct the mass bumps of the heavy charged Higgs boson and the associated top quark. For this purpose, we propose a hybrid-R reconstruction method which utilizes the top tagging technique, a jet substructure technique developed for highly boosted massive particles. By using the full hadronic mode of $p p \to H^{\pm} t \to t tb$ as a test field, we find that this method can greatly reduce the combinatorics in the full reconstruction and can successfully reduce background events down to a controlled level. The  sensitivity of LHC to the heavy charged Higgs boson with two $b$ taggings is studied and a $9.5\sigma$ significance can be achieved when $m_{H^\pm} =1 \textrm{TeV}$.}

\begin{document}

\section{Introduction}
Probing the Higgs sector and unraveling the miracle of electroweak symmetry breaking (EWSB) is one of the most important goals at the Large Hadron Collider (LHC). In the Standard Model (SM), the Higgs sector includes a physical neutral Higgs boson and three would-be Goldstone bosons, while in most of extensions of the SM more than one physical Higgs bosons are predicted \cite{HiggsHunter}. For example, two Higgs doublet model (THDM) extends the SM by adding one additional $SU(2)$ doublet and results in a rich phenomena in Higgs sector by boasting two CP-even Higgs bosons, $h$ and $H$, one CP-odd Higgs boson $A$, and a charged Higgs boson $H^{\pm}$. The discovery of neutral Higgs bosons and the measurement to their properties could undoubtedly help people understand the EWSB mechanism, while the confirmation/exclusion of the existence of a charged Higgs boson at the LHC can pinpoint down the whole Higgs sector and further distinguish models.

Experiments have invested quite an effort into the search of charged Higgs bosons. There is a direct limit of $M_{H^{\pm}}>78.6$ GeV from the LEP searches by its exclusive decay of $H^{\pm}\rightarrow \tau\nu$ and $H^{\pm}\rightarrow cs$ \cite{LEP}. At hadron colliders, the search approaches for a charged Higgs boson differ in term of its mass range. For the low mass range $m_{H^{\pm}} < m_t$, the signal for a charged Higgs boson is from top quark decay $t\rightarrow H^+b$ followed by the decay $H^+ \rightarrow \bar{\tau}\nu$. At the Tevatron, due to the effects of phase space suppression for a heavy charged Higgs boson production, the search is mainly focused on the low mass range $m_{H^{\pm}}<m_t$. Currently there is no charged Higgs signal detected from top decay nor obvious deviation from $Wtb$ vertex measurement, which can put a constraint to THDM on the small and large $\tan \beta$ regions for a charged Higgs boson mass up to $\sim 160$ GeV \cite{Tevatron}. However, due to its large collision energy of the LHC, the search for a heavy charged Higgs boson can be feasible via the $gb\rightarrow tH^-$ production up to a large mass range \cite{TDR}. For the large mass range $m_{H^{\pm}}>m_t$, the signal is from the main production process the $gb$ fusion ($gb\rightarrow tH^-$) followed by its decays. When a charged Higgs boson is heavy enough, its main decay mode can be $H^-\rightarrow t\bar{b}$ while other modes (say $H^-\rightarrow \tau\bar{\nu}$) are small and can be neglected.


When the heavy charged Higgs boson is around $1 \textrm{TeV}$ or higher, the top quark from its decay can be highly boosted and the detectors of LHC may not resolve all jets from its decay when the angle separation parameter is fixed to a specific value (say $R=0.5$ in anti-kt jet algorithm). Motivated by new technique of ``jet substructure" \cite{Jetography,VermilionThesis,boost2010,SpannowskyRev} developed for highly boosted massive particles, we propose a ``hybrid-R reconstruction method" and investigate the full hadronical decay channel of the heavy charged Higgs production. Previous studies on this process focused on its  semi-leptonic decay, i.e. $tH^{\pm}\rightarrow t{\bar t} b \rightarrow jjbl\nu bb$ and mainly depend on triple-$b$-tagging \cite{GunionHC,BargerHC,treepptbH} or four-$b$-tagging \cite{4b} to suppress the background. To our best knowledge, this is the first attempt to the full hadronically decaying channel for process $pp\rightarrow tH^{\pm}\rightarrow t{\bar t} b$. Similar to its semi-leptonic decay, the full reconstruction of its full hadronic decay mode may be difficult due to its large combinatorics and the large backgrounds from $t\bar{t} + \textrm{jets} $ and QCD multi-jets processes \cite{TDR}. Furthermore, the $b$ jet from the boosted top may be difficult to be tagged. Therefore, we will use the top tagging and the $b$ tagging for other isolated $b$ jets as well as the full reconstructed objects in the final state to suppress the background.

In order to identify objects in the signal, we reconstruct all objects (i.e. two W bosons, two top quarks, and a charged Higgs boson) in the signal. In our hybrid-R reconstruction method, we first use a larger cone size for the top tagger to identify a highly boosted top. Then we utilize the fact that the $b$ jet from the charged Higgs boson can be reliably resolved by detectors with a smaller cone size and should be the most energetic jet to reduce the combinatorics in the full reconstruction. The remaining task is to identify the associated top quark which is always produced near threshold of the process where all jets from this top quark decay can be resolved by the smaller cone. In short, our hybrid-R reconstruction utilizes the advantages of both a larger cone (for a top tagging) and a smaller cone (for other resolvable jets). And it is found that this method can work quite well when the charged Higgs boson mass is above $800 \textrm{GeV}$.

To be more precise to define our hybrid-R reconstruction method, we firstly adopt a large cone size in the top tagging method in order to capture the highly boosted top (John Hopkins top-tagger\footnote{It is a top tagging method proposed by John Hopkins group and we will introduce it in section \ref{sec:toptagging}} \cite{JHToptagging:2008} in Cambridge-Aachen algorithm with $R$ optimized for each $m_{H^\pm}$ value). Then we adopt a smaller radius $R$ (anti-$k_t$ with $R$=0.4) to resolve jets in the event, which is efficient to reconstruct the associated top with a small transverse momentum and the isolated $b$ jet from the heavy charged Higgs boson decay. We introduce a few crucial cuts to suppress background and adopt the neural network and boosted decision tree to optimize cuts.
It is found that, with the help of John Hopkins top-tagger and hybrid-R reconstruction method, the full-hadronic decay of a heavy charged Higgs boson at the LHC ($\sqrt{s}=14 $TeV) is reachable. We also study the sensitivity of LHC to a heavy charged Higgs boson varying from $0.8 \textrm{TeV}$ to $1.5 \textrm{TeV}$ and find that LHC can cover this heavy mass region when the integrated luminosity is assumed to be 100 fb$^{-1}$.

The remainder of this paper is organized as follows.  In \Sec{sec:jet}, we firstly give a brief review of jet algorithms and jet substructure. In \Sec{sec:HC}, we introduce the ``hybrid-R reconstruction method". And then we employ the ``hybrid-R reconstruction method" armed with top tagging method to study the process $pp \rightarrow tH^{\pm} \rightarrow t{\bar t} b$ with both tops in hadronic decay mode in \Sec{sec:results}. Our conclusions follow in \Sec{sec:conclusions}.

\section{Brief Review of Jet Algorithms and Jet Substructure}
\label{sec:jet}
\subsection{Jet Algorithms and Jet Substructure}
A Jet algorithm is to reconstruct a jet from the collimated spray of hadrons and to make comparison between data and theoretical prediction. Jet algorithms can be casted into two categories: the cone-based algorithms and the sequential recombination algorithms. The former one relies on the intuitive idea to put a cone along the dominant direction of the energy flow. The later one repeatedly recombines the closest pair of particles which look like undoing the QCD branching. In this paper, we will focus on sequential recombination algorithms including the $k_t$ \cite{Kt1,Kt2}, the Cambridge-Aachen \cite{CA1,CA2} and the anti-$k_t$ \cite{AKT}. For a recent review of jet algorithms and its application at hadron collider, we refer to Ref. \cite{Jetography}.

Sequential recombination jet algorithms begin with a list of four-momenta of pseudo-jets which can be reconstructed from the track detectors and calorimeters, both electromagnetic and hadronic ones, and recursively combine pairs of momenta into jets. The combination rules can be described as :

\begin{enumerate}
\item Calculate the distance $d_{ij}$ between all pairs of pseudo-jets and the beam distance $d_{iB}$ for each pseudo-jet.
\item Find the minimum in the list of $d_{ij}$ and $d_{iB}$.
\item If the smallest entry is $d_{ij}$, recombine i and j into a new particle and then return to step 1.
\item Otherwise, if it is $d_{iB}$, define it as a final-state jet, and remove it from the original pseudo-jet list. Return to step 1.
\item Iterate this process until the original list is empty, i.e., all pseudo-jets have been clustered to jets.
\end{enumerate}

The $d_{ij}$ and $d_{iB}$ in most popular sequential recombination algorithms for use at hadron collider can be parameterized as follows~\cite{AKT}:
\bea
  \label{eq:genkt-dist-R}
    d_{ij} &=& \min(p_{ti}^{2p}, p_{tj}^{2p}) \frac{\Delta R_{ij}^2}{R^2}\,,\qquad\quad
    \Delta R_{ij}^2 = (y_i - y_j)^2 + (\phi_i - \phi_j)^2
    \,,\\
    d_{iB} &=& p_{ti}^{2p}\,,
\eea
where $p$ is a parameter that is $1$ for the $k_t$ algorithm, -1 for the anti-$k_t$ algorithm and $0$
for the Cambridge-Aachen algorithm (CA algorithm). The $k_t$ algorithm clusters soft particles first, while the anti-$k_t$ algorithm recombines hard particles first. And the CA algorithm clusters by angle ranking. There are a few comments in order:
\begin{itemize}
\item There is no standard jet algorithms. The jet algorithm and the parameters could be chosen flexibly.
\item All these algorithms are infrared-safe and collinear-safe.
\item The anti-$k_t$ algorithm gets circular jets area which act much like an idealized cone-based algorithm, while the $k_t$ algorithm and CA algorithm get irregular jets area. It is one of reasons why experimental groups favor the anti-$k_t$ from the viewpoint of detector geometry.
\item The $k_t$ and CA algorithms incline to combine the harder pseudo-jets or pseudo-jets with a large separation in the later, so it is expected that the combination in last step is the reverse of the main QCD branching or the 2-body decay of a massive particle which can be easily used in jet substructure study by iteratively tracing backwards the clustering history. However, the recombination history in anti-$k_t$ algorithm couldn't be utilized
 to related the branching and to get the subjets.

 \end{itemize}

The study of jet substructure has received considerable attention in recent years. At the LHC, the massive electroweak particles, i.e., $W$, $Z$, Higgs and top quarks, are always produced at a high transverse momentum ($P_t \gg M_Z$). In this case, their decay products tend to become collimated due to the large Lorentz boost factor. For highly boosted massive particles with hadronic decay, it is expected that a large cone size $R$ could capture all the collimated decay products in a ``fat jet". And the mass of this ``fat jet" is one indicator of its origin \cite{
Skiba:2007fw,Holdom:2007ap, Agashe:2007ki}. However, high energy QCD jets can also generate large masses. In order to suppress QCD background to an acceptable level, one have to resolve more physical information in order to pick out the ``fat jet" originated from the decay of highly boosted massive particles. One way is to anatomize the substructure and to identify the subjets in a fat jet to reconstruct the delicate kinematical variables.

Along the line of this thought, the jet substructure technique has a boosting development in recent years including early studies in theory \cite{Seymour:1993mx,Butterworth:2002tt,BDRS,JHToptagging:2008,BoosttopWang,Nojiri:2008ir,tth:09,Trimming:2009,Pruning1,Pruning:2009PRD} and in experiment \cite{YSplitter,AtlasHWHZ,AtlasTop08,AtlasTop09,CMS-boosted-top}. For recent reviews on jet substructure, see Refs.
\cite{Jetography,VermilionThesis,boost2010,SpannowskyRev}.
The earliest study of jet substructure is Ref. \cite{Seymour:1993mx} in the context of a search for a heavy Higgs boson decaying to WW. More recently, the BDRS algorithm \cite{BDRS} developed a technique using the mass-drop and the filtering to transform the high-$P_t$ $WH, ZH ( H \rightarrow b\bar{b} )$ channel into one of the best channels for discovery of Standard Model Higgs with small mass at the LHC.
The filtering \cite{BDRS}, i.e., reclustering the fat jet of Higgs with refined small cone size $R_{filt}$, is supposed to further reduce the degradation of resolution on jets caused by underlying events.
Besides filtering, some new techniques including ``trimming", ``pruning", ``variable R" are also developed to refine the jet resolution
\cite{Trimming:2009,Pruning1,Pruning:2009PRD,VariableR}. Application of the BDRS method to Higgs search in the MSSM context can be found at \cite{arXiv:0912.4731,arXiv:1006.1656}.

Below, we focus on the top tagging algorithms.


\subsection{Top Tagging}
\label{sec:toptagging}
Top quarks play an important role in understanding electroweak symmetry breaking and searching for new physics. Unlike the case at Tevatron where most of top quarks are produced near the threshold, at the LHC many boosted top quarks can be created.

In the conventional method, top quarks are reconstructed by recombining three daughters, i.e , three isolated jets in hadronic decay mode or one isolated lepton, one isolated jet and large missing energy in leptonic decay mode. And $b$-tagging always is used to suppress the large QCD background.

However, this conventional method confronts some problems for the top with high-$P_t$.
\begin{itemize}
\item The decay products become collimated and it is difficult to find three isolated daughters to reconstruct the top when a cone size is fixed to a specific value like $R=0.5$ in the anti-$k_t$ algorithm.
\item The $b$-tagging efficiency drops significantly as the $P_t$ increases because the tracks become too close  to clearly identify the secondary vertex from the B decays in the silicon detector.
\item For its leptonic decay, it is a task to determine the missing energy in the hadron collision environment.
\end{itemize}
So, the jet substructure technique used to identify boosted tops from its hadronic decay have been developed in recent years. As summarized in Ref. \cite{Jetography}, many discriminating variables are used on subjets for top tagging,
such as $d_{ij}$-types variables \cite{AtlasTop08,AtlasTop09,Nsubjettiness},
$P_t$-ratio type variables \cite{JHToptagging:2008,BoosttopWang,tth:09,Pruning1,Pruning:2009PRD,Nsubjettiness},
event-shape variables \cite{Sterman:2008,Topjets:2008}, and constraints on a $W$-subjet mass\cite{JHToptagging:2008,BoosttopWang}
as well as other subjets correlation variables ( a helicity angle $\theta_t$ \cite{JHToptagging:2008} ). In addition, the HEPTopTagger\cite{tth:09,HEPTopTagger}
is designed for moderately boosted tops and some techniques are developed for leptonic top tagger \cite{LepTopTagger}.

Here, we adopt the top tagging method for the highly boosted top in hadronic decay proposed in Ref. \cite{JHToptagging:2008}. We will dub it as ``JH top-tagger" for simplicity and use it in later analysis.
As well-known, a boosted top from its hadronic decay looks like a fat jet with three hard cores.
Similar to the BDRS jet substructure method for boosted Higgs, the JH top-tagger firstly uses a large cone to cluster the event in order to capture all the decay products and relevant radiations of a boosted top and then declusters the top jet to find three subjets.

To resolve a fat top jet into the relevant hard substructure, the following recursive procedure are applied in JH top-tagger\cite{JHToptagging:2008}:
 \begin{enumerate}
\item Four momenta of particles is clustered with a large cone size $R$ ( CA algorithm are used in original paper)
\item The last cluster is undone to get two objects $j_1$ and $j_2$. If the $P_t$ ratio of the softer jet $j_2$ over the original jet $j$ is too small, i.e., $P_{t_{j_2}}/P_{t_j}< \delta _p$ , throw the softer $j_2$ and go on to decluster on the harder one.
\item The declustering step is repeated until two separated hard objects are found. If any criterion below are satisfied, the event is failed: 1) both objects
are softer than $\delta _p$ (2) the two objects are too close, $ \Delta \eta + \Delta \phi < \delta_r $ (3)the original jet is considered irreducible.
\item Declustering repeatedly on these two subjets will result in 2,3, or 4 hard objects.
\item The case with 3 or 4 subjets are kept. And then require the these subjets that the total mass should be near $ m_t$, the mass of two subjets should be in the $ m_W$ window, $W$ helicity angle $\theta_t$ should be consistent with a top decay\footnote{The helicity angle $\theta_t$ is defined
as the angle, measured in the rest frame of the reconstructed $W$, between
the reconstructed top's flight direction and one of the $W$ decay products. The lower $P_t$
subjet in the lab frame are chosen to set the angle. }.
\end{enumerate}
The parameters involved in the method can be optimized event by event \cite{JHToptagging:2008}. In our study, these parameters are fixed as below:
\bea
\label{eq:para}
\delta_p=0.19 \qquad \delta_r=0.1 \qquad |\theta_t|<0.65\,\,. \\ \nonumber
\eea
An interesting application of the Higgs tagging method as well as the top quark tagging method for the heavy top partner search can be found at \cite{arXiv:1012.2866}.

\section{Heavy Charged Higgs in Full Hadronic Decay Mode}
\label{sec:HC}
\subsection{Heavy Charged Higgs}
For a heavy charged Higgs boson ($m_{H^{\pm}}>m_t$), the main production process is the $gb$ fusion process $gb\rightarrow tH^-$. And the LHC can probe the charged Higgs boson in a wide mass range up to TeV scale. Additional productions including $W^{\pm} H^{\mp}$ \cite{ppHW}, $\Phi H^{\pm}$ ($\Phi=A,\ h,\ H$) \cite{ppPhiH} and  $H^{+} H^{-}$ \cite{ppHH} have also been studied but the cross sections are generally small in these modes.
For the $H^{\pm}$ with $m_{H^{\pm}}>m_t + m_b$, the decay $H^- \rightarrow \bar{t}b$ is the main decay mode  due to the large Yukawa coupling. Although the decay mode $H^- \rightarrow \tau\bar{\nu}$ is an interesting one, its small branching ratio makes it difficult to probe a heavy charged Higgs boson.
The early studies on the search for a charged Higgs on $pp\to tH^{\pm} $ production can be found in Ref \cite{GunionHC,BargerHC,treepptbH,4b}. The search in this channel are
also carried out by both ATLAS and CMS collaborations. These collaborations mainly investigate
$gb\rightarrow tH^-\rightarrow t{\bar t} b \rightarrow bqq\tau(had) \nu bb$ and the semi-leptonical channel
$gb\rightarrow tH^-\rightarrow t{\bar t} b \rightarrow bqqbl \nu b$.
For search in $\tau$ channel $gb\rightarrow tH^-\rightarrow t{\bar t} b \rightarrow bqq\tau(had) \nu b$, the branching ration of $H^{\pm} \rightarrow \tau\nu$ is small especially for higher $H^{\pm}$ mass. The main backgrounds to this signal are $t\bar{t}$ events, $W$+jets and QCD multi-jet events \cite{TDR}.
For semi-leptonically decaying channel of $H^{\pm}$ production, i.e., $tH^{\pm}\rightarrow t{\bar t} b \rightarrow jjbl\nu bb$, the main backgrounds come from $t \bar{t} + \textrm{jets}$, $t \bar{t} + b \bar{b}(QCD)$, and $t \bar{t} + b \bar{b} (EW)$ \cite{TDR}. These studies indicate the search for a heavy charged Higgs boson from this channel has to rely on triple-$b$-tagging \cite{GunionHC,BargerHC,treepptbH} or four-$b$-tagging \cite{4b} in order to suppress the background.

Armed with the jet substructure method to tag highly boosted top quarks, we investigate the full hadronic decay mode for process $pp \rightarrow gb \rightarrow t_2H^{-} \to t_2\bar{t}_{1}b_3$\footnote{We also include $pp \rightarrow gb \rightarrow \bar{t}_{2}H^{+} \to \bar{t}_{2}t_1\bar{b}_3$ in our calculation}. To our best knowledge, this is the first attempt for such a study in literature. For convenience, we here call the top quark and the $b$ quark from the charged Higgs boson decay as $t_1$ and $b_3$, respectively, and the associated top quark as $t_2$ as well.
We attempt to reconstruct all objects in the final states. For this purpose, we resort to the top tagging method and a ``hybrid-R reconstruction method" which can help to overcome the combinatorics problem in the full reconstruction while maintaining a remarkable suppression to backgrounds (especially QCD background).

Since we only concentrate on the study of the interaction of the heavy charged Higgs with top and bottom quarks, we simplify the Lagrangian as follows:
\bea
{\cal L} &=& {\cal L}_{SM} + \partial_{\mu} H^{\pm} \partial^{\mu} H^{\mp} + m_{H^\pm}^2 H^{\pm} H^{\mp} \nonumber \\
&&+ H^{\pm} {\bar t} (Y_L P_L + Y_R P_R) b +   h.c.\,\,.
\eea
The couplings $Y_L$ and $Y_R$ are treated as free parameters in the model-independent analysis. In two Higgs doublet model, for simplicity, we assume the branching ratio of charged Higgs into $tb$ is $100\%$ and take $Y_L = Y_R = 1$.



\subsection{Hybrid-R Reconstruction Method for Full Hadronic Decay Mode}
\label{sec:hybrid}
For the full hadronically decaying of process $pp \rightarrow gb \rightarrow t_2 H^{-} \to t_2\bar{t}_{1}b_3$, the main backgrounds are $t\bar{t}$ and QCD multi-jets events. We focus on a heavy charged Higgs boson with mass around 1 TeV or so.

In such a mass region, the daughter $t_1$ from the heavy charged Higgs boson decay is always highly boosted. And a larger cone size $R$ is expected to capture all the hadronic decay products of $t_1$ in a fat jet. However, the associated $t_2$ in the $pp\to t_2H^{\pm}$ process is produced near its threshold and its decay products generally can not form a fat jet. So the large cone size used for $t_1$ tagging fails to catch the $t_2$ and only a small cone size in a jet algorithm can resolve all its three jets. Then the traditional top reconstruction method by recombining three jets is expected to be efficient to tag $t_2$.

In order to reconstruct all objects in the process $pp \rightarrow gb \rightarrow t_2 H^{-} \to t_2\bar{t}_{1}b_3$ and to exploit the advantages of a larger and a smaller cone size jet algorithms, we propose a ``hybrid-R reconstruction method".

The five steps we used in our ``hybrid-R reconstruction method" can be described as follows:
\begin{enumerate}
\item  At the preselection level, we adopt the CA jet algorithm with a larger cone size parameter and use a top tagger method to capture a highly boosted top and preselect events. The JH top-tagger is employed in this work. The jets at this step form the jet set $J_0$.
\item  After capturing a top jet labeled as $t_1$, we re-cluster the pseudo-jets in the event with a smaller cone size parameter (the anti-$k_t$ algorithm with $R=0.4$ is chosen as an example) and then get a new jet list $L_0$.
\item  If a small-size jet is within the larger cone of the direction of the tagged top jet, we remove it from the jets list for the $b_3$ and $t_2$ reconstruction. The rest of jets form a new jet list $L_1$.
\item We identify the most energetic jet in the list $L_1$ as a $b_3$ jet. This choice also implies that the tagged top $t_1$ and the identified $b_3$ fly back-to-back. This is due to the observation that the tagged top in the larger cone size by the top-tagger method contains subjets which are massive and energetic and the leading two objects with large transverse momentum in the smaller cone size always fly back to back. So we don't impose a cut on $\Delta \phi$ or $\Delta R$ here. The unused jets form the list $L_2$.
\item We require there are at lest 3 jets in the jet list $L_2$ in an event. Then we introduce a $\chi^2$ method to tag the second top quark in the signal.
Here, $\chi^2 = \frac{|m(j_1,j_2) - m_W^{\textrm{PDG}}|^2}{\sigma_W^2} + \frac{ |m(j_1,j_2,j_3) - m_t^{\textrm{PDG}}|^2 }{\sigma_t^2}$ where $\sigma_W$ and $\sigma_t$ are
equal to $10$ GeV and $15$ GeV, respectively, which is to account for the detector resolution capability. The $m_W^{\textrm{PDG}}$ and $m_t^{\textrm{PDG}}$ are taken as
$80.5$ GeV and $173$ GeV, respectively. The combination which minimize the $\chi^2$ is chosen as the right reconstruction.
\end{enumerate}


In step 2 we recluster a selected event with a smaller cone size, and such a reclustering has also been used to reduce the degradation of the resolution on a fat jet from highly boosted massive particle in filtering \cite{BDRS} and some other grooming techniques, and also can be used as preselection rule \cite{tth:09}, and can be used to separate top jets from QCD jets in the hollow cone method \cite{Hollowtop} as well. But it should be stressed that, in our hybrid-R reconstruction method, the reclustering and removing the smaller jets within the radius of the tagged top is to catch the highly boosted top and to reconstruct successfully the other massive particle with a smaller $P_t$ (Here it is the $t_2$).

Another remarkable point is that any top tagging algorithm for highly boosted top can be used in the first step although the JH top-tagger is used as an example.

\begin{figure}[!htb]
\begin{center}
\includegraphics[width=1.0\columnwidth]{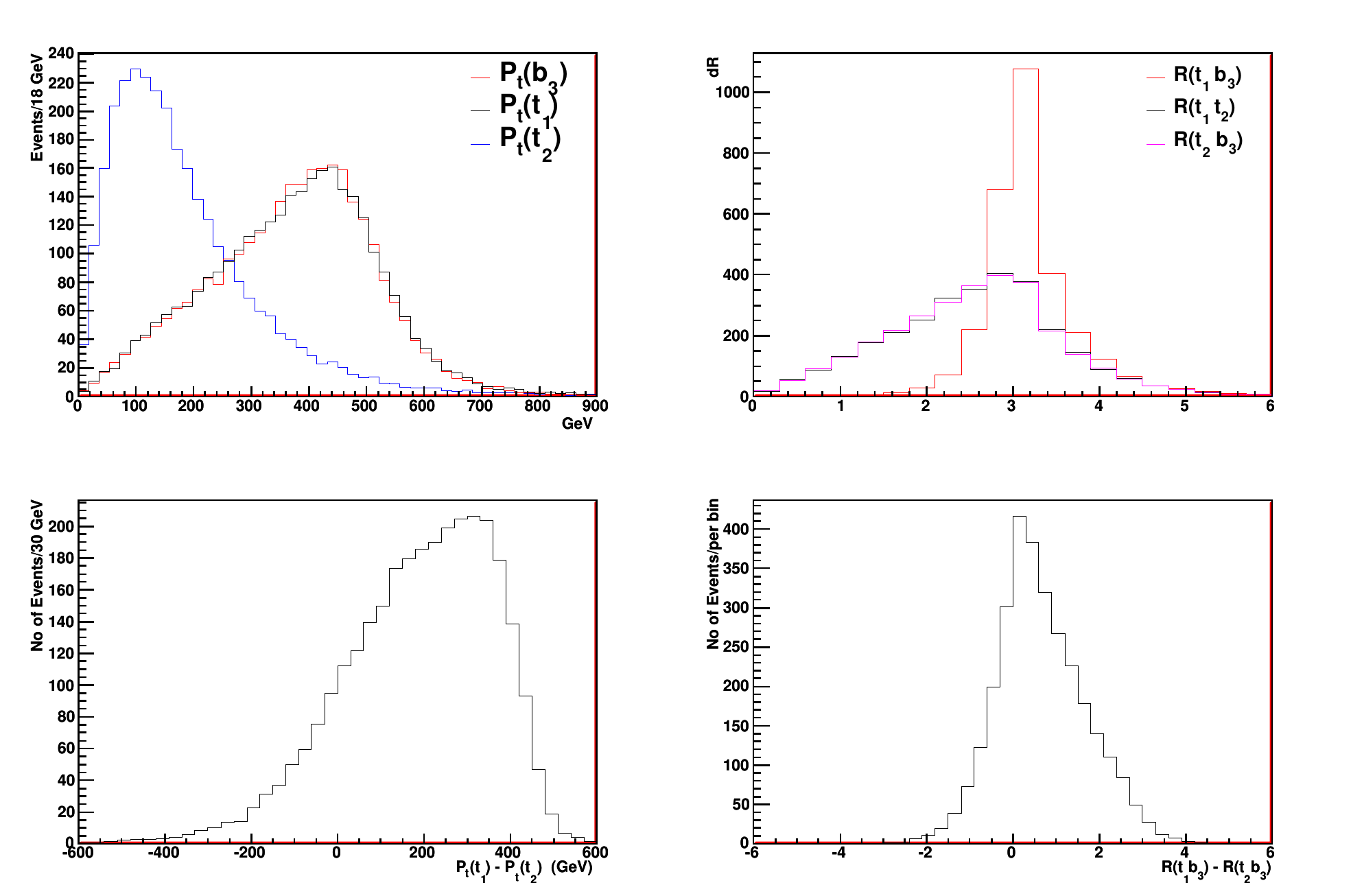}
\caption{The transverse momenta and the angle separation are exposed at parton level with $m_{H^{\pm}}=1.0 \textrm{TeV}$. The transverse momentum difference between $t_1$  and $t_2$ and the angle separation difference between $R(t_1 b_3)$ and $R(t_2 b_3)$ are also shown.
\label{partonfig}}
\end{center}
\end{figure}

Before carrying on the detailed analysis, we would like to show some features of the signal which support our hybrid-R reconstruction method. At parton level, in Fig. \ref{partonfig}, we show the distribution of transverse momenta of $t_2$, $t_1$, and $b_3$, and the angle separation among two of $t_1$, $t_2$ and $b_3$ in the case of $M_{H^{\pm}}=1$ TeV. The distributions of $P_t(t_1)$ and $P_t(b_3)$ are similar and yield an important clue for $P_t(t_1)$ and $P_t(b_3)$. We find that $P_t(t_1)=P_t(b_3) \geq \frac{3}{10} m_{H^{\pm}}$ can work and so we will use it in our simple cut method which will be introduced in the following section. We also show the distribution of $P_t(t_1) - P_t(t_2)$ and $R(t_1 b_3) - R(t_2 b_3)$, which provide important hints to determine the right combination. As shown in Fig. \ref{partonfig}, one can find the fact that in most of phase space the transverse momentum of $t_1$ is larger than $t_2$. Meanwhile, the angle separation between $t_1$ and $b_3$ demonstrates that these two objects indeed fly back-to-back. And in most of cases, the angle separation between $t_1$ and $b_3$ is larger than that between $t_2$ and $b_3$. At parton level, it is also found that the $\Delta \phi(t_1, b_3)$ is always close to $\pi$, which may be useful in the full reconstruction.

\begin{figure}[!htb]
\begin{center}
\includegraphics[width=1.0\columnwidth]{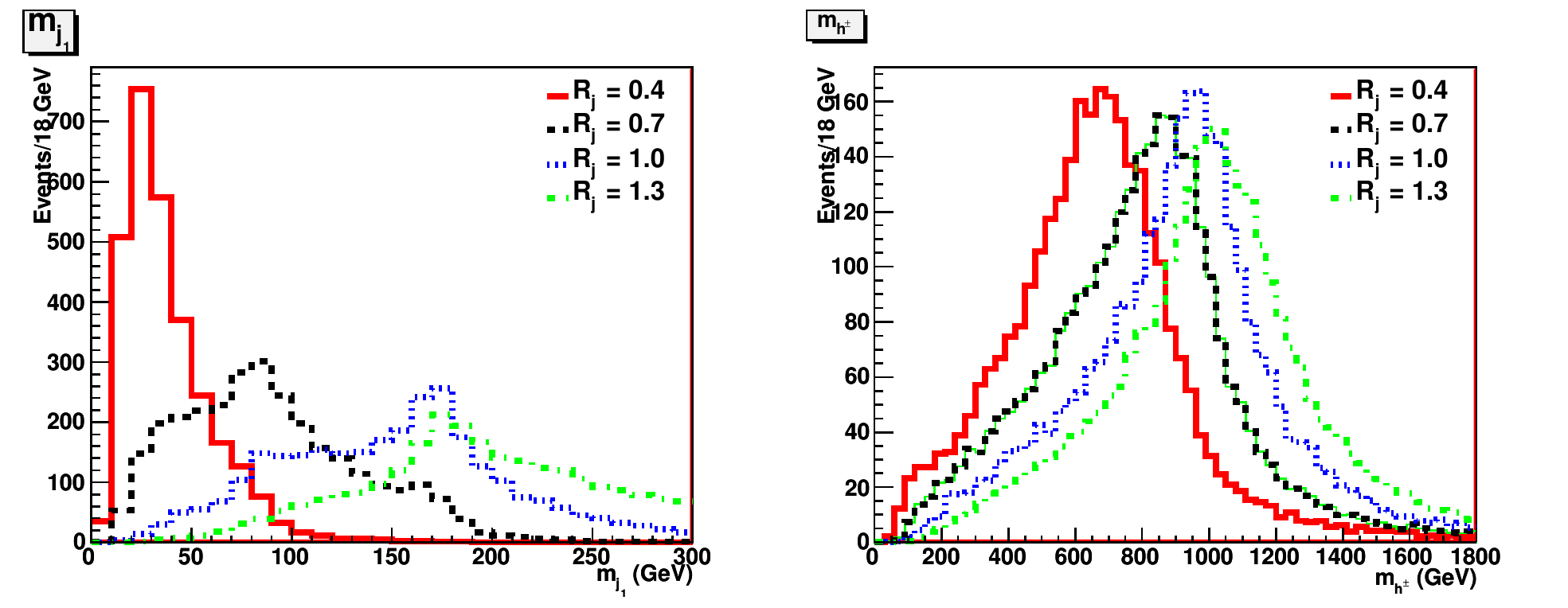}
\caption{The mass distribution of the leading massive jet and the reconstructed charged Higgs boson mass are shown with the case $m_{H^{\pm}}=1 \textrm{TeV}$. The dependence on the angle separation parameter $R$ in the anti-$k_t$ algorithm is also exposed. To be more realistic, the detector effects have been taken into account by using PGS4.
\label{massfig}}
\end{center}
\end{figure}

At the fast detector simulation level by using PGS4 \cite{PGS}, as demonstrated in the left panel of Fig. \ref{massfig}, we show the mass distribution of the leading massive jet, which is assumed to be a $t_1$ jet in our hybrid-R reconstruction method. Then from the rest of jets, we pick a jet with the largest transverse momentum as the $b_3$ jet. After picking out these two jets, we examine the invariant mass of the combined Lorentz vector from them, which is also displayed in the right side of Fig. \ref{massfig}. We examine how these two types of mass distribution vary to the change of angle separation parameter $R$ in the anti-$k_t$ algorithm.

In Fig. \ref{massfig}, we can read out that when $R=0.4$, neither a $W$ boson nor a top quark can be captured in a single massive jet significantly. When $R=0.7$, a considerable fraction of W boson can be captured in one single massive jet while a small fraction of $t$ quark can be captured. When $R=1.0$ and $R=1.3$, quite a fraction of top quark can be captured by a single massive jet.

Meanwhile, when $R$ is large enough, the simple reconstruction method by using the most-massive jet and the most-energetic jet can even reconstruct the heavy charged Higgs boson mass bump, as demonstrated in Fig. \ref{massfig}. Both parton level and fast detector simulation level analysis demonstrate that our event reconstruction method captures the most important feature of the signal.

In our Monte Carlo study, the signal events are generated by Madgraph/MadEvent \cite{Madgraph} and background events are produced by Alpgen \cite{Alpgen}. These events are fed to DECAY to generate full hadronic decay final states and pass to PYTHIA \cite{Pythia} to simulate showering, fragmentation, initial state radiation, final state radiation, and multi-interaction as well. After that, we use fastjet \cite{Fastjet} and SpartyJet\cite{Spartyjet} to perform jet clustering and analysis.

Now, we examine the top tagging efficiency of JH top-tagger for different values of the cone size parameter $R$ in both $k_t$ and CA algorithms by using the SpartyJet package. The results are tabulated in Table \ref{table1.0tev} and Table \ref{table1.5tev}, which are two cases study for $m_{H^\pm}= 1.0 \textrm{TeV}$ and $m_{H^\pm}= 1.5 \textrm{TeV}$, respectively. As shown in Table \ref{table1.0tev} and Table \ref{table1.5tev}, the tagging efficiency of the CA algorithm is better than that of the $k_t$ algorithm in a larger R, especially for a larger value of $m_{H^{\pm}}$. Another remarkable point is that the efficiency of the CA algorithm does not depend upon the value of $R$ as that of the $k_t$ algorithm does in large $R$ range. Due to these feature, in later analysis, we will use the top-tagger in CA algorithm with optimized cone sizes for different $m_{H^{\pm}}$.

\begin{table}[th]
\begin{center}%
\begin{tabular}
[c]{|c|c|c|c|c|c|c|c|c|}\hline
        $R$ &  $0.7$& $0.8$ & $0.9$&  $1.0$ & $1.1$ & $1.2$ & $1.3$ & $1.4$ \\\hline
$k_t$ algorithm &  $16\%$ & $22\%$ &  $25\%$ & $28\%$ & $29\%$ & $29\%$& $28\%$& $27\%$\\ \hline
CA algorithm    &  $13\%$ & $19\%$ &  $24\%$ & $27\%$ & $29\%$ & $30\%$ & $30\%$ & $30\%$ \\ \hline
\end{tabular}
\end{center}
\caption{The efficiency of top tagger in $k_t$ and CA algorithms along with the variation of angle separation parameter $R$ are tabulated. The charged Higgs boson mass is assumed to be 1.0 TeV.}%
\label{table1.0tev}%
\end{table}

\begin{table}[th]
\begin{center}%
\begin{tabular}
[c]{|c|c|c|c|c|c|c|c|c|}\hline
        $R$ &  $0.7$& $0.8$ & $0.9$&  $1.0$ & $1.1$ & $1.2$ & $1.3$ & $1.4$ \\\hline
$k_t$ algorithm &  $25\%$ & $27\%$ &  $28\%$ & $28\%$ & $27\%$ & $26\%$& $24\%$& $22\%$\\ \hline
CA algorithm    &  $23\%$ & $27\%$ &  $29\%$ & $30\%$ & $30\%$ & $30\%$ & $30\%$ & $29\%$ \\ \hline
\end{tabular}
\end{center}
\caption{The efficiency of top tagger in $k_t$ and CA algorithms along with the variation of angle separation parameter $R$ are tabulated. The charged Higgs boson mass is assumed to be 1.5 TeV.}%
\label{table1.5tev}%
\end{table}

This behavior can be understood by the clustering sequence of jet algorithms. In the CA algorithm, the clustering is performed by angle ranking and the softer drop method, i.e., $\delta _p$ cut in the top-tagger will
efficiently remove the softer branching, underlying events and pile-up with a large separation to the hard subjets. Therefore it is successful to find the refined radius for each hard core in a top jet. However, in the $k_t$ algorithm, the clustering recombines the hardest and largest separated particles in the last stage of recombination, the softer pseudojets have been merged ahead and the $\delta _p$ cut actually don't work as it does in the CA algorithm.
Additionally, in the $k_t$ algorithm, a jet has a larger area and a stronger $P_t$ dependence than that in the CA algorithm \cite{jetarea}.

These features also explain why the efficiency of top-tagger in the CA algorithm with a smaller $R$ is smaller than that in the $k_t$ algorithm. In a smaller cone size, the CA algorithm can not capture all jets from the top quark decay while the $k_t$ algorithm is better since it can bring more protojets and get a larger mass for its larger jet area and no softer drop.

It is also remarkable that the configuration of the $b$ quark and the $W$ boson from $t_1$ decay flying near to the direction perpendicular to the boost direction of $t_1$ has larger probability to form a top jet. While the configuration either the $b$ or the $W$ boson flying parallel to the boost direction can not form a top jet. The parameters of top tagging given in Eq. (\ref{eq:para}) also affect the tagging efficiency. This explains why there are more than $70\%$ or so in the signal can not be successfully tagged as a top jet.

\section{Analysis Results}
\label{sec:results}

In this section, we present our detailed analysis. One obvious question is whether signals can be recorded at detectors of LHC. Below we use the triggering system of ATLAS detector as an example to examine whether the signals can be recorded.

At the level-1 triggering, signals can be selected out by the jet triggering from heavy charged Higgs boson decay which have large $P_t$. The mutli-jet trigger should also help since there are more than $3$ jets in the signals. Furthermore, the scalar sum of transverse energy, which should be close to the heavy charged Higgs boson mass, can also help to pick out the signals. At the level-2 triggering, the tagged $b$ jets can be helpful to record signals. At the event filter level, the top tagger algorithm can be implemented to select signals.

At the offline analysis level, we introduce a set of cuts and compare their efficiencies between the signal and backgrounds. Multivariate analysis methods, the neural network and the boosted decision tree, are also used to improve the significance.

\subsection{Signal and Backgrounds }

\begin{figure}[h]
\centerline{
\epsfxsize=7 cm \epsfysize=5 cm \epsfbox{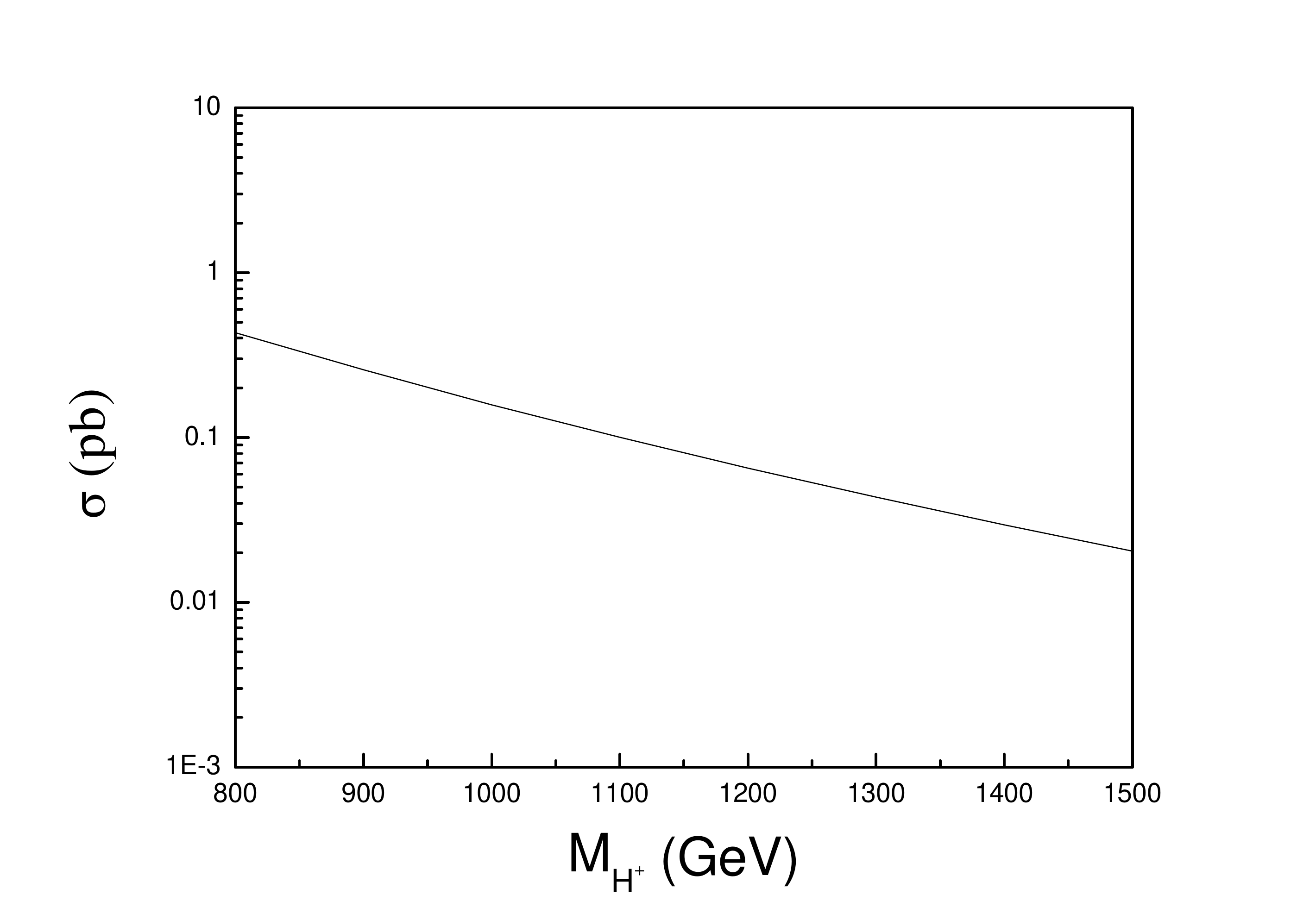}
\hspace*{0.2cm}
\epsfxsize=7 cm \epsfysize=5 cm \epsfbox{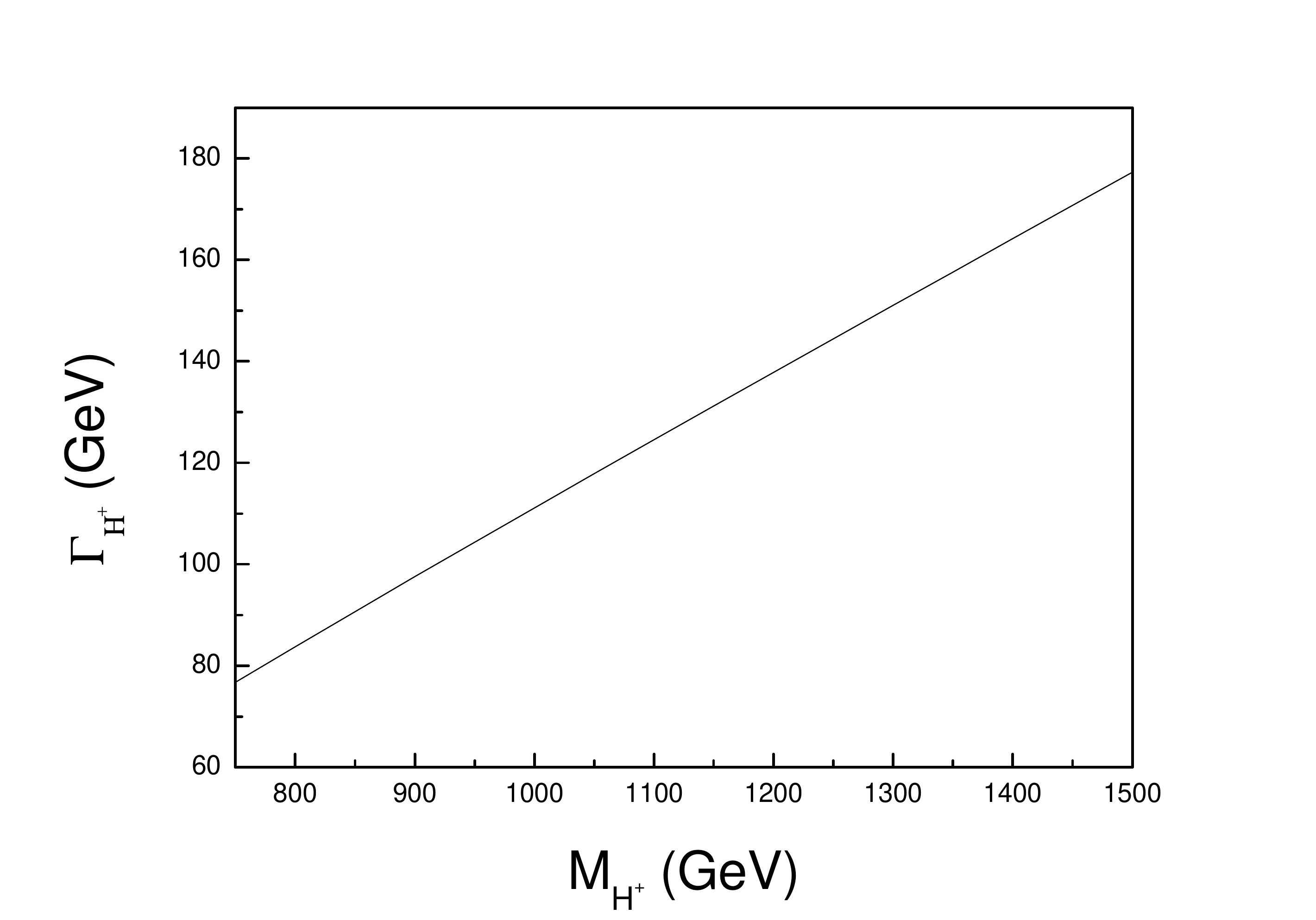}}
\vskip -0.05cm \hskip 3.4 cm \textbf{( a ) } \hskip 6.2 cm \textbf{( b )}
\caption{(a) The cross section as a function of  $M_{H^{\pm}}$ at the LHC with $\sqrt{s}=14$ TeV is plotted. (b) The width as a function of $M_{H^{\pm}}$ is shown.}
\label{figmt}
\end{figure}



\begin{figure}[htbp]
\begin{center}
\includegraphics[width=1.0\columnwidth]{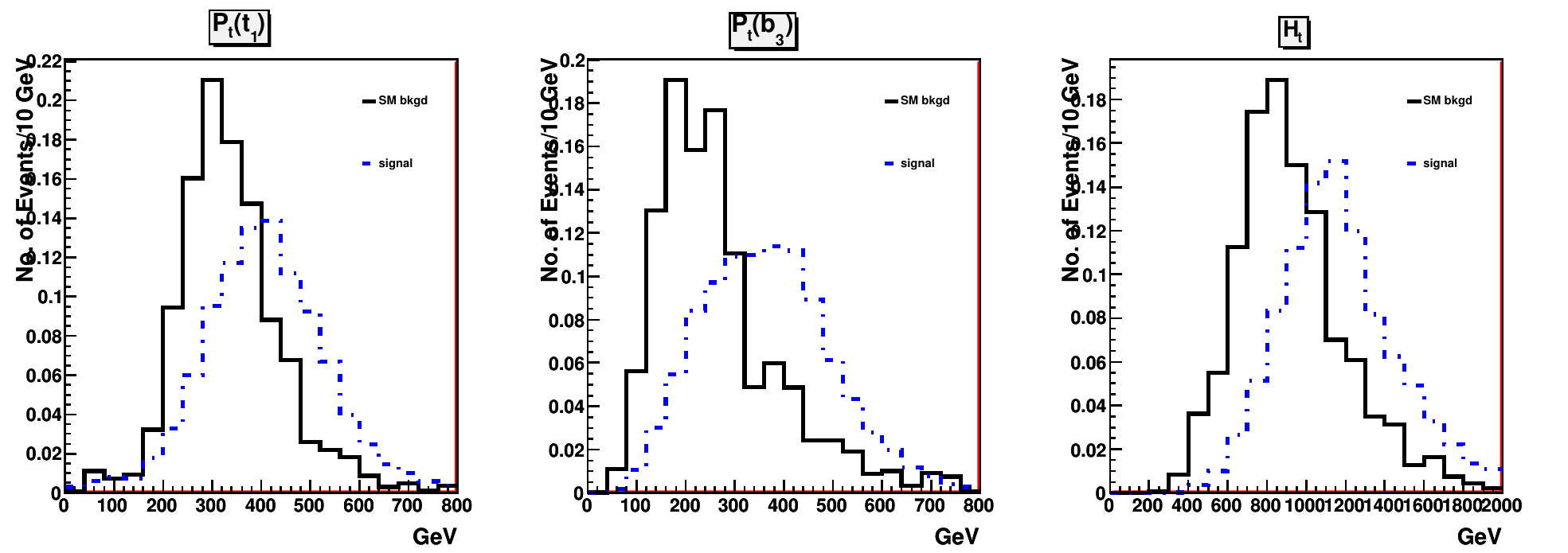}
\caption{For the case $m_{H^\pm}=1 \textrm{TeV}$, the normalized distributions of signal and backgrounds for kinematic observables, $P_t(t_1)$, $P_t(b_3)$, and $H_t$, are demonstrated, respectively.}
\label{fig-rc-1}
\end{center}
\end{figure}

\begin{figure}[htbp]
\begin{center}
\includegraphics[width=1.0\columnwidth]{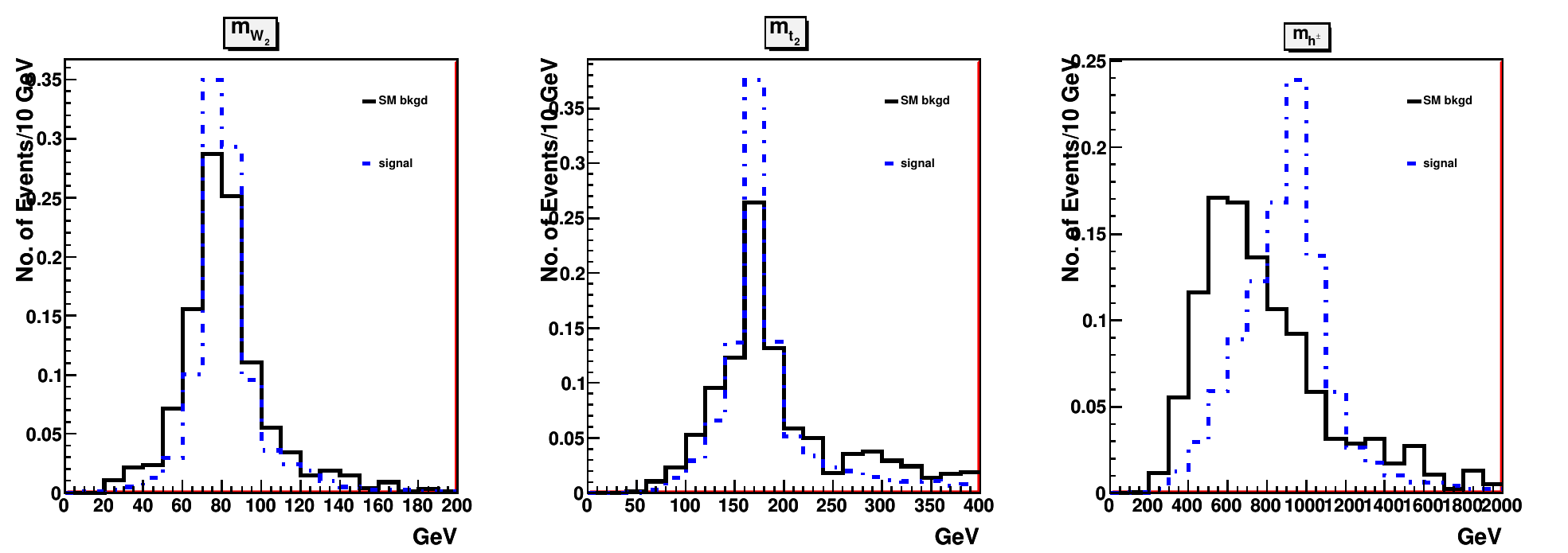}
\caption{For the case $m_{H^\pm}=1 \textrm{TeV}$, the normalized distributions of both signal and background for reconstructed mass peaks, $m_{W^2}$, $m_{t^2}$, and $m_{H^\pm}$ are shown.}
\label{fig-rc-2}
\end{center}
\end{figure}

The cross section of the process $pp \rightarrow gb \rightarrow t_2H^{-} \to t_2\bar{t}_{1}b_3$ and the decay width of heavy charged Higgs boson are shown in Fig. \ref{figmt}. The cross section is about $0.2 \textrm{pb}$ for $m_{H^\pm}=0.8 \textrm{TeV}$, which drops down to $0.02 \textrm{pb}$ for $m_{H^\pm}=1.5 \textrm{TeV}$. While the decay width is about $75 \textrm{GeV}$ for $m_{H^\pm}=0.8 \textrm{TeV}$, which increases to $170 \textrm{GeV}$ for $m_{H^\pm}=1.5 \textrm{TeV}$. It is found that the decay width can be approximated by $\Gamma_{m_{H^\pm}} \approx 10\% \times m_{h^\pm}$. After using the PGS4 to simulate the detector effect, it is also found that the effect of decay width to the mass bump of $H^\pm$ is roughly comparable to the detector effect, which means the mass bump of the charged Higgs boson is not too broad to be detectable from the detectors of the LHC.

Since the cross section of SM background processes is huge, it is necessary to use the $b$-tagging (assuming $b$-tagging efficiency 60\% and mis-tagging efficiency for g and quarks 1\%) to suppress the SM background events.
We demand that the most energetic isolated jet and one extra $b$ jet in the jet list $L_0$ must be tagged as $b$ jets.
The most energetic isolated jet must be a tagged $b$ jet can suppress the $t{\bar t}+\textrm{jets}$ background by a factor of $100$ since the isolated jet in this process is mainly from gluon radiation.

Below we introduce a few crucial kinematic observables. One of the most characteristic features of the signal is that the $t_1$ and $b_3$ from the decay of heavy charged Higgs boson are very energetic and fly in the small $\eta$ region. Therefore, the top-tagger is efficient to capture the highly boosted $t_1$ coming from the decay of the heavy charged Higgs since the large mass of $m_{H^{\pm}}$ will transform to the high $P_t$ of the top quark $t_1$. For $t\bar{t}j$ background, there is only a fraction of events could provide a high $P_t$ top jet passing the JH top-tagger. However, it is difficult to find a massive jet has the same substructure as the highly boosted top in QCD multi-jet events. Meanwhile, the $H_T$ observable, i.e, the scalar $P_t$ sum of final states, should be a good discriminant for signal and background separation.

After tagging a highly boosted $t_1$, for the signal the $b_3$ jet generally should be the most energetic jet in the rest of jet list $L_1$. It is also expected that the distribution of transverse momentum of $t_1$ and $b_3$ should be good for signal and background separation. Once both $t_1$ and $b_3$ are identified, the mass bump of the heavy charged Higgs boson can be reconstructed. Then the only combinatorics is from the reconstruction of the second top quark, $t_2$, which can be reconstructed from the remain jet list $L_2$. While it is difficult for backgrounds to reconstruct such a top quark especially for QCD multi-jet. So it is obvious that the reconstruction mass bumps from $t_2$ including $m_{W_2}$ and $m_{t_2}$ are important to separate the signal and backgrounds.

In Figs. \ref{fig-rc-1}-\ref{fig-rc-2}, distributions of kinematic variables of events which pass the preselection rules are presented. As shown in Fig. \ref{fig-rc-1}, the distribution of $P_t(t_1)$, $P_t(b_3)$, and $H_T$ for both signal and background events are shown. In Fig. \ref{fig-rc-2}, the mass bump of $m_{W_2}$, $m_{t_2}$, and $m_{H^\pm}$ are shown.

According to these kinematic features, a set of preselection cuts are chosen and the comparison of the cuts efficiencies between the signal and backgrounds are shown in Table \ref{table:comparison1.0} and Table \ref{table:comparison1.5}. These two tables correspond to the case for heavy charged Higgs with mass 1 TeV and 1.5 TeV at the LHC ( $\sqrt{s}$=14 TeV), respectively. Due to the preselection rules introduced in our reconstruction method, the more relevant QCD multijet background are generated by requiring the leading four jets with $P_{t,min}>100 \textrm{GeV}$.

Some of the preselection rules, which have been encoded in our reconstruction method, are listed as
\begin{itemize}
\item Detector acceptance cuts are chosen as $P_t(j)>20$ GeV, $|\eta|(j)<2.5$.
\item A cut on the number of jets in the jet list $L_0$, $n_j \geq 6$.
\item A cut on the $H_T$, $H_T>\frac{7}{10} M_{H^{\pm}}$, and the centrality, $C>0.3$.
\item A cut on the leading energetic jet in the jet list $L_0$, $E(j_1)>\frac{7}{20} M_{H^{\pm}}$.
\item A cut on the second leading energetic jet in the jet list $L_0$, $E(j_1)>\frac{1}{4} M_{H^{\pm}}$.
\end{itemize}

At the data analysis level, for the sake of comparison with the multivariate analysis methods, we introduce a simple cut method consisting of several important cuts after the full reconstruction, which are listed below
\begin{itemize}
\item A large $P_t$ requirement on $t_1$ jet captured by the top-tagger and on the supposed $b_{3}$ in hybrid-R reconstruction method, i.e. $P_t(b_3)>\frac{3}{10} M_{H^{\pm}}$ and $P_t(t_1)>\frac{3}{10} M_{H^{\pm}}$.
\item Cuts are chosen for the mass bumps in the $t_2$ reconstruction, i.e., $|m_{W_2} - m_{W}^{\textrm{PDG}} | < 20$ GeV and  $|m_{t_2} - m_{t}^{\textrm{PDG}}| < 30 GeV$.
\item The charged Higgs boson mass window is chosen as  $|m_{H^\pm} - m^{\textrm{assumed}}_{H^\pm}| < 200$ GeV.
\end{itemize}
The efficiency of each cut is also shown in Tables \ref{table:comparison1.0} and \ref{table:comparison1.5}. The effects of $b$ taggings, which can help to reduce combinatorics and suppress QCD background events, are also shown. Here the $b$ tagging requires the leading $P_t$ jet must be tagged, where we assume $b$-tagging efficiency as $0.6$ and the mistagging efficiency as $0.01$.


\begin{table}[th]
\begin{center}%
\begin{tabular}
[c]{|c|c|c|c|}\hline
 & signal &   $t \bar{t} + \textrm{jets}$ & QCD \\
 & $m_{H^\pm}=1$ TeV  &  & $n_j \geq 4$ \\\hline
 Cross Section x Br (pb)  &  $0.053$ & $553$  & $9186.8$ \\ \hline \hline
After JH tagger \& $H_T>400 \textrm{GeV}$ &  $29\%$ & $7 \%$ &   $4.8 \times 10^{-3}$ \\ \hline
Two $b$-taggings & $10\%$ & $4.2 \times 10^{-4}$ & $4.8 \times 10^{-7}$ \\ \hline \hline
The number of jets in jet list $L_0$, $n_j \geq 6$  &  $10\%$ &   $3.7 \times 10^{-4}$ & $3.4\times10^{-7}$ \\ \hline
$H_T>700 \textrm{GeV}$ \& $C > 0.3$ &  $9.6\%$ &   $2.1 \times 10^{-4}$ & $3.0 \times 10^{-7}$ \\ \hline
The leading jet  $E(j_1)>350 \textrm{GeV}$&  $8.9\%$ & $1.6 \times 10^{-4}$ &   $2.5 \times 10^{-7}$ \\ \hline
The second leading jet $E(j_2)>250 \textrm{GeV}$&  $7.9\%$ & $1.3 \times 10^{-4}$ &   $2.2 \times 10^{-7}$ \\ \hline \hline
$P_t(b_3)>300 $ GeV \& $P_t(t_1)>300$ GeV  &  $5.3\%$ &   $ 4.5 \times 10^{-5}$ & $7.7 \times 10^{-8}$ \\ \hline
$|m_{W_2} - m_{W}^{\textrm{PDG}} | < 20$ GeV &  $3.1\%$ & $2.6 \times 10^{-5} $ &   $2.3 \times 10^{-8}$ \\
\& $|m_{t_2} - m_{t}^{\textrm{PDG}}| < 30$ GeV& & & \\\hline
$|m_{H^\pm} - m^{\textrm{assumed}}_{H^\pm}| < 200 $ GeV &  $2.5\%$ & $1.5 \times10^{-5}$ &   $1.5 \times 10^{-8}$ \\ \hline \hline
Events in 100 fb$^{-1}$ & $133$ & $830$ & $13.8$ \\ \hline
\end{tabular}
\end{center}
\caption{The efficiencies of both signal and background before and after using JH top tagger are displayed. The collision energy is assumed to be 14 TeV. The JH top tagger and CA algorithm are used with the angle separation parameter $R=1.1$.}%
\label{table:comparison1.0}%
\end{table}

\begin{table}[th]
\begin{center}%
\begin{tabular}
[c]{|c|c|c|c|c|}\hline
 & signal &  $t \bar{t} + \textrm{jets} $& QCD  \\
 &  $m_{H^\pm}=1.5$ TeV &  & $n_j \geq 4$ \\\hline
 Cross Section x Br (pb)  &   $0.007$ &$553$ & $9186.8$ \\ \hline \hline
After JH tagger \&  $H_T>400 \textrm{GeV}$&  $29\%$ & $3 \%$ &   $3.3 \times10^{-3}$ \\ \hline
Two $b$-taggings & $10\%$ & $1.8\times 10^{-4}$ & $3.3 \times 10^{-7}$ \\ \hline \hline
The number of jets in jet list $L_0$, $n_j \geq 6$  &  $9.3\%$ &   $1.7 \times 10^{-4}$ & $2.5\times10^{-7}$ \\ \hline
$H_T>1050 \textrm{GeV}$\& $C > 0.3$ &  $8.3\%$ &   $5.5 \times 10^{-5}$ & $8.9 \times 10^{-8}$ \\ \hline
The leading jet  $E(j_1)>525 \textrm{GeV}$&  $7.7\%$ & $4.4 \times 10^{-5}$ &   $7.3 \times 10^{-8}$ \\ \hline
The second leading jet $E(j_2)>375 \textrm{GeV}$&  $6.8\%$ & $3.4\times 10^{-5}$ &   $5.4\times 10^{-8}$ \\ \hline \hline
$P_t(b_3)> 450$ GeV \& $P_t(t_1)> 450$ GeV  &  $4.4\%$ &   $9.2 \times 10^{-6}$ & $8.5 \times 10^{-9}$ \\ \hline
$|m_{W_2} - m_{W}^{\textrm{PDG}} | < 20$ GeV&  $2.5\%$ & $4.6 \times 10^{-6}$ &   $4.3 \times 10^{-9}$ \\
$|m_{t_2} - m_{t}^{\textrm{PDG}}| < 30$ GeV & & & \\\hline
$|m_{H^\pm} - m^{\textrm{assumed}}_{H^\pm}| < 200 $ GeV &  $1.8\%$ & $1.9 \times 10^{-6}$ &   $1.8 \times 10^{-10}$ \\ \hline \hline
Events in 100 fb$^{-1}$ & $13$ & $105$ & $0.2$ \\ \hline
\end{tabular}
\end{center}
\caption{The number of events of both signal and background before and after using JH top tagger are displayed. The collision energy is assumed to be 14 TeV. The JH top tagger and CA algorithm are used with the angle separation parameter $R=0.9$.}%
\label{table:comparison1.5}%
\end{table}

As shown in Table \ref{table:comparison1.0}, the top tagging efficiency is 30\% for signal while the mistagging efficiency for QCD background are at 0.1\% level. For $t\bar{t}j$ background, there is a small fraction about $7\%$ where one of the top quark can be boosted. As for $M_{H^{\pm}}=1.5$ TeV, the top tagging efficiency and the mistagging efficiency are similar. The comparison of the cuts efficiencies for the signal and backgrounds are shown step by step. After imposing the requirement of successful reconstruction of $t_2$, the background have been considerably rejected. Furthermore, if a mass window cut on the reconstructed charged Higgs boson is imposed, the background is significantly suppressed by a factor 3 or so for the SM background while the signal is retained in $50\%$ (for $M_{H^{\pm}}=1$TeV). It is of interest to note that the cuts used in above steps can suppress the background by a factor $10$ ($100$) for $m_{H^\pm}=1 \textrm{TeV}$ ($m_{H^\pm}=1.5 \textrm{TeV}$) at least. This suppression factor increases with the increase of charged Higgs boson mass.

We would like to make a comment from Table \ref{table:comparison1.0} and Table \ref{table:comparison1.5}. These two Tables provide crucial evidences to use the top tagging substructure technique and $b$ taggings.
Without the top tagger and $b$ taggings to select the most relevant events, the computing and analysis time for background events will increase drastically, which might be difficult to make the full hadronic decay mode realistic. Furthermore, if the standard top reconstruction methods are used for $t_1$ reconstruction, we have to confront with the combinatorics problem. Even the most energetic top quark $t_1$ can be successfully reconstructed, it is found that the dominant background $t{\bar t}$+jets can not be sufficiently suppressed. Similarly, without the top tagging, the mistagging rate from QCD background can reach to a few percent. For the sake of comparison, we have considered a full reconstruction without using the top tagger. We have considered events with number of jets $n_j \geq 7$ and have tried a biased $\chi^2$ method, where $\chi^2$ is defined as
\bea
\chi^2 & =\frac{|m_{W_{1}}-m_W^{\textrm{PDG}}|^2}{\sigma_W^2} + \frac{|m_{W_{2}}-m_W^{\textrm{PDG}}|^2}{\sigma_W^2} \nonumber  + \frac{|m_{t_{1}}-m_t^{\textrm{PDG}}|^2}{\sigma_t^2} + \frac{|m_{t_{2}}-m_t^{\textrm{PDG}}|^2}{\sigma_t^2}  \\ & + \frac{|\delta \phi(t_1,b_3) - \pi|^2}{\sigma_\phi^2}\,.
\eea
In order to reduce combinatorics and to save processing time, among the leading two jets of each event, we assume one must be $b_3$ and the other must from $t_1$ decay. The last term in Eq. (4.1) is to capture the configuration with $t_1$ and $b_3$ from the heavy charged Higgs boson decay, and we have used $\sigma_\phi = 0.2$. Then the mass of the heavy charged Higgs boson can be reconstructed. Without the help of TMVA methods, for $m_{H^\pm}=1$ TeV ($m_{H^\pm}=1.5$ TeV) case, we find the significance can not be better than $4$ ($1.5$) due to a large number of background events and the deterioration of wrong reconstruction of signal events. While when top tagger is used, we can greatly reduce combinatorics and analysis time and maintaining a better significance. Therefore the top tagging methods adopted here are indeed necessary. It is also remarkable that the $b$ taggings are also crucial to suppress the SM background (especially the QCD background) to a computable and controllable level. Without $b$ taggings, the significance can not be better than $1.2$($0.5$). 

\begin{figure}[!htb]
\begin{center}
\includegraphics[width=1.0\columnwidth]{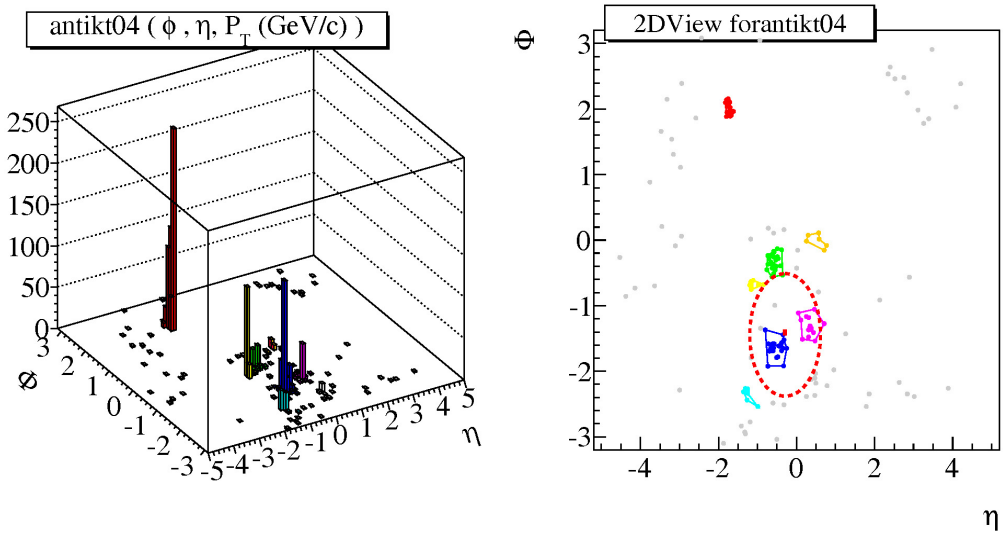}
\caption{The lego plot of $P_t$ and 2-Dimension plot of $\eta-\Phi$ for a typical signal event clustering with anti-$k_t$ $R=0.4$ for illustration. And the direction of the top jet tagged by JH top-tagger and the top area with $R=1.0$ are marked with a red circle.
\label{fig:eventplot}}
\end{center}
\end{figure}

In Fig. \ref{fig:eventplot}, we also show the $P_t$ lego plot and 2-Dimension $\eta - \phi$ diagram of a typical signal event (anti-$k_t$ $R$=0.4) passing our selection rules. The direction of the top jet tagged by JH top-tagger and the top area with $R=1.0$ are marked with a red circle. In this example, the traditional recombination method fails because only two jets can be found with the jet algorithm parameter $R$=0.4 within the tagged top jet. However, such a signal event can be efficiently selected out by our hybrid-R reconstruction method equipped with the top tagging algorithm. Furthermore, our method is efficient to reconstruct the other top because we reduce substantially the combinatorics after subtracting the jets within the tagged top and removing the $b_3$ jet. For a naive comparison, the combinatorial number in the traditional recombination method is $C_{7}^{3}=35$ to reconstruct a top in one event with seven jets. In our method, the combinatorial number is $C_{4}^3=4$.

\subsection{Multivariate Analysis and Search Bounds}
In this subsection, we explore the multivariate analysis (TMVA) to optimize the cuts and improve the sensitivity. Both of the neural network discriminant analysis and the boosted decision tree discriminant analysis are considered. TMVA methods have been broadly used in the top quark precision measurements by experimental collaborations \cite{Pleier:2008ig,:2007qf,Abazov:2006yb} and  W-jet substructure analysis \cite{Cui:2010km}, as well as new physics search studies, like Higgs boson search \cite{Gallicchio:2010dq} and heavy quarks search \cite{AguilarSaavedra:2009es,Holdom:2010fr,Holdom:2011uv}. The boost decision tree method has been adopted in the light charged Higgs boson search \cite{Ali:2011qf}.

\begin{figure}[htbp]
\begin{center}
\includegraphics[width=1.0\columnwidth]{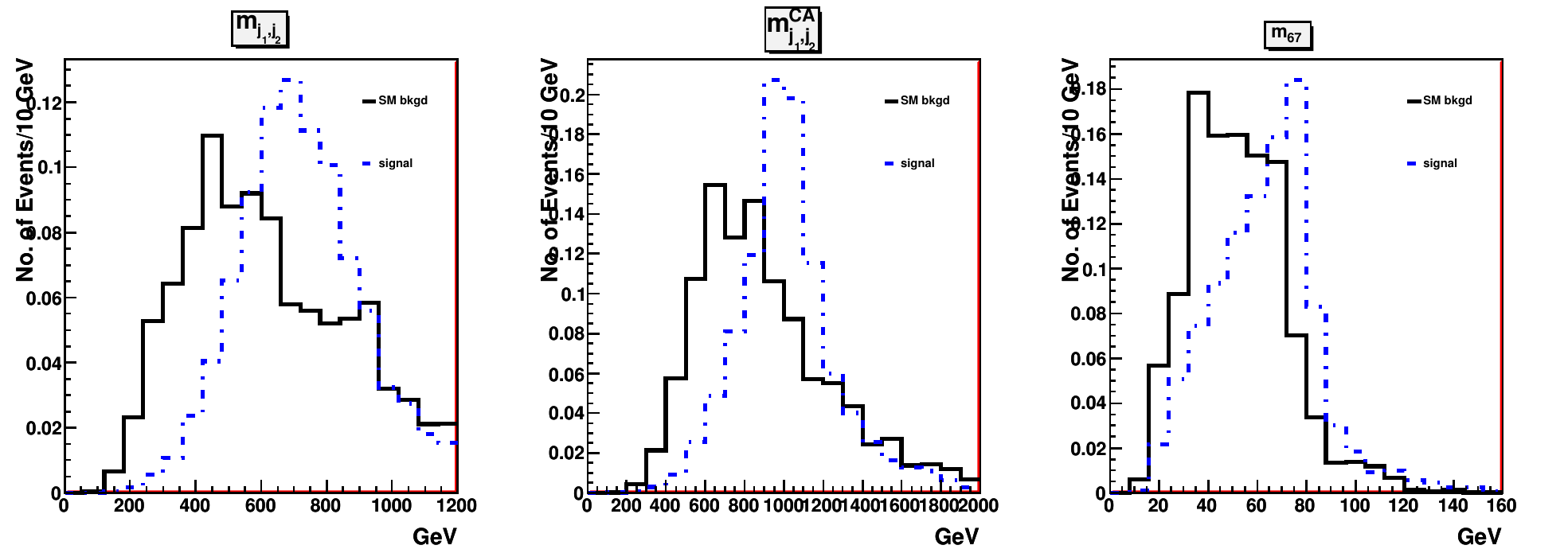}
\caption{For the case $m_{H^\pm}=1 \textrm{TeV}$, the normalized distributions of both signal and background for a few kinematic observables, $m_{j_1 j_2}$, $m_{j_1 j_2}^{CA}$, and $m_{67}$ are shown.}
\label{fig-rc-3}
\end{center}
\end{figure}
In Fig. (\ref{fig-rc-3}), three extra kinematic observables are shown. Except the kinematic variables introduced in Section 4.1, we also find several extra kinematic observables helpful to separate signal and background.
\begin{itemize}
\item The observable $m_{j_1,j_2}$ is the maximal invariant mass of a pair of jets in the jet list $L_0$.
\item The observable $m_{j_1 j_2}^{CA}$ is the invariant mass of the leading two jets in the jet list $J_0$. This observable can form a bump for signals near the mass of heavy charged Higgs boson.
\item The observable $m_{67}$ is the minimal invariant mass of a pair of jets for the leading $7$ jets in the jet list $L_0$. The signals tend to condense near the mass of W boson.
\item The maximal value of invariant mass of leading three jets in the jet list $L_0$, which is labeled as $m^{\textrm{x}}_{123}$, is found to be concentrated near the mass of heavy charged Higgs boson for signal.
\item The minimal value of invariant mass of three jets among the leading 7 jets in the jet list $L_0$, which is labeled as $m_{567}$, is found to be concentrated near the mass of top quark for signal.
\item The invariant mass of the leading 7 jets in the jet list $L_0$ is found to be useful for signal and background discrimination.
\end{itemize}

\begin{figure}[!htb]
\begin{center}
\includegraphics[width=1.0\columnwidth]{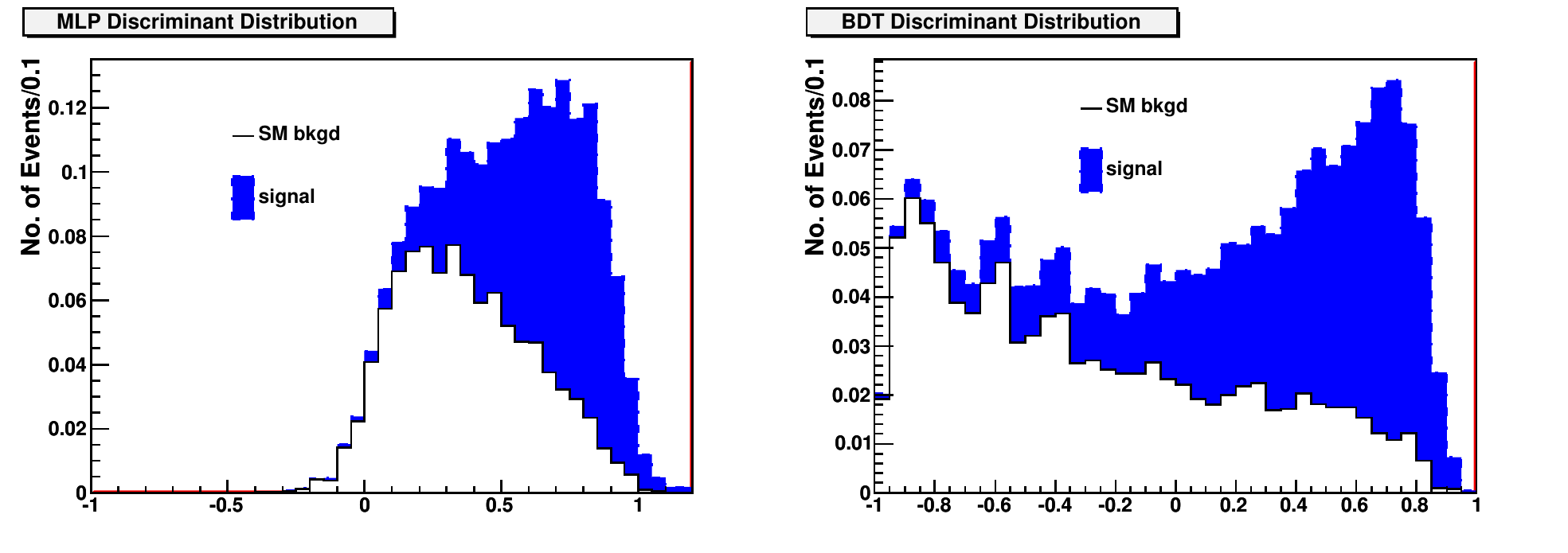}
\caption{The discriminant distributions of neural network analysis and boosted decision tree analysis are demonstrated, where both signal and background is normalized to one.}
\label{discrimination}
\end{center}
\end{figure}

\begin{figure}[!htb]
\begin{center}
\includegraphics[width=1.0\columnwidth]{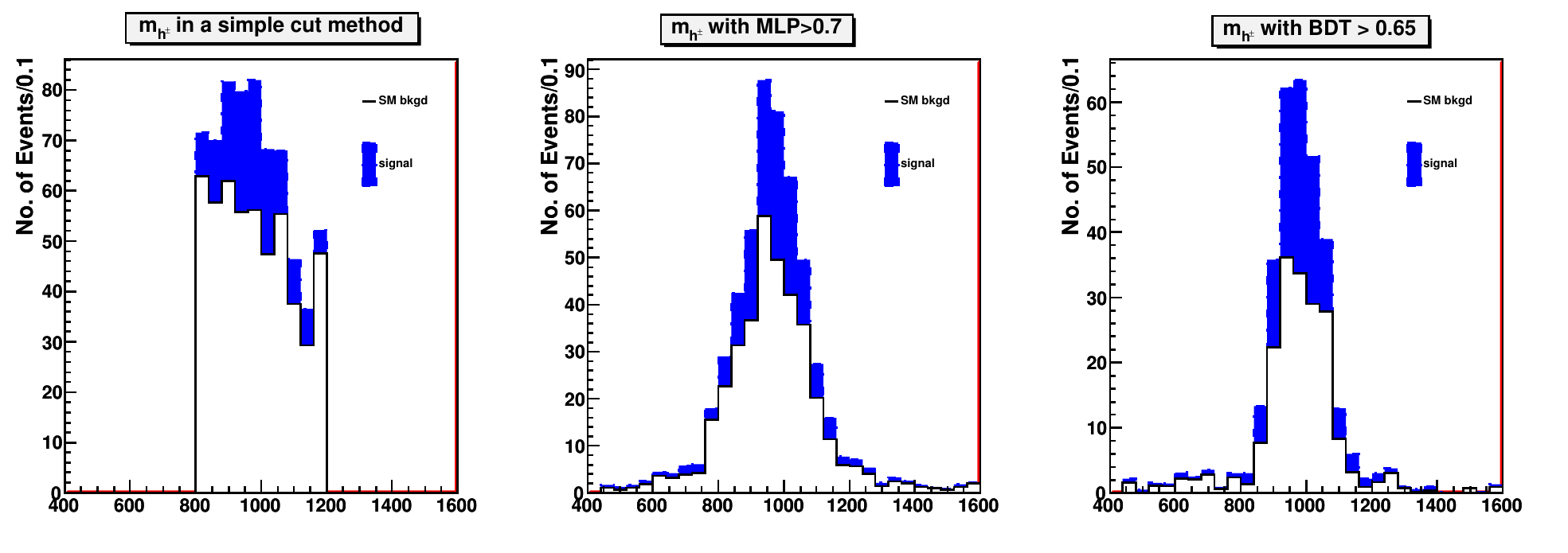}
\caption{The reconstructed mass bump of charged Higgs boson in three analysis methods
are demonstrated, where both signal and background have been normalized to one before analysis.}
\label{massbumps}
\end{center}
\end{figure}

The discrimination distributions of multivariate analysis are demonstrated in Fig. \ref{discrimination}. As it is shown, multivariate analysis methods (Neural Network and Boosted Decision Tree) using the input kinematic observables indeed work well to separate the signal from background events. And the efficiencies of these two TMVA methods are also listed in the Tables \ref{table:TMVA1.0} and \ref{table:TMVA1.5}, as a comparison to that of the simple cut method used above. As shown in the Tables \ref{table:TMVA1.0} and \ref{table:TMVA1.5}, both of the neural network and boosted decision tree methods can improve the significance by a factor $100\%$ or more. For instance, for the charged Higgs with mass 1 TeV, both TMVA methods can improve the significance by gaining more signals at the price for a little bit more background events. This improvement is also demonstrated in the mass window of the charged Higgs boson, as demonstrated in Fig. \ref{massbumps}.


\begin{table}[th]
\begin{center}%
\begin{tabular}
[c]{|c|c|c|c|c|}\hline
 & signal &  $t \bar{t} + \textrm{jets} $& QCD  \\
 &  $m_{H^\pm}=1.0$ TeV &  & $n_j \geq 4$ \\\hline
After simple cuts& $2.5\%$ & $1.5\times 10^{-5}$ & $1.5 \times 10^{-8}$ \\ \hline
After NN cut ($NN>0.6$)&  $5.5\%$& $2.0\times 10^{-5}$ & $3.0 \times 10^{-8}$ \\ \hline
After BDT cut ($BDT>0.5$) &  $5.7\%$& $2.1\times 10^{-5}$ & $3.2 \times 10^{-8}$ \\ \hline
\end{tabular}
\end{center}
\caption{Comparison among the efficiencies of the neural network, boosted decision tree methods and hybrid-R reconstruction method by step-by-step cuts for $M_{H^{\pm}}=1$ TeV.}
\label{table:TMVA1.0}%
\end{table}

\begin{table}[th]
\begin{center}%
\begin{tabular}
[c]{|c|c|c|c|c|}\hline
 & signal &  $t \bar{t} + \textrm{jets} $& QCD\\
 &  $m_{H^\pm}=1.5$ TeV &  &  $n_j \geq 4$ \\\hline
After simple cuts&  $1.8\%$ & $1.9 \times 10^{-6}$ &   $1.8\times 10^{-10}$ \\ \hline
After NN cut ($NN>0.7$)&  $4.0\%$& $2.0\times 10^{-6}$ & $3.6 \times 10^{-10}$ \\ \hline
After BDT cut ($BDT>0.5$) &  $4.5\%$& $2.1\times 10^{-6}$ & $2.4 \times 10^{-10}$ \\ \hline
\end{tabular}
\end{center}
\caption{Comparison among the efficiencies of the neural network, boosted decision tree methods and hybrid-R reconstruction method by step-by-step cuts for $M_{H^{\pm}}=1.5 $ TeV.}%
\label{table:TMVA1.5}%
\end{table}

\begin{table}[th]
\begin{center}%
\begin{tabular}
[c]{|c|c|c|c|c|c|c|c|c|}\hline
$m_{H^\pm}$ (TeV) & $0.8$ &  $0.9$  & $1$ & $1.1$ & $1.2$ &$1.3$ &$1.4$ &$1.5$ \\ \hline
$\sigma$ (fb) & 324 & 192& 118& 75 & 49& 33& 22 & 15 \\ \hline \hline
$\frac{S}{\sqrt{B}}$ (with two $b$ taggings \& TMVA) & 19.4 & 13.3 & 9.5 & 7.0 & 5.1 & 4.0 & 3.1 & 2.6 \\ \hline
lower bound on $\sigma(fb)$ & 33 & 29 & 25 & 21  & 19  & 16  & 14 &  11 \\ \hline \hline
$\frac{S}{\sqrt{B}}$ (with two $b$ taggings without TMVA) & 9.3 & 6.5 & 4.6 & 3.4 & 2.5 & 1.9 &1.5 & 1.3  \\ \hline
lower bound on $\sigma(fb)$ & 70  & 59 &  51 &  44 & 39  & 35  & 29 &  23 \\ \hline
\end{tabular}
\end{center}
\caption{The significance and sensitivity of LHC for heavy charged Higgs boson production are shown, where we assume the total integrated luminosity as 100 fb$^{-1}$. For the sake of comparison, the results with and without TMVA are provided. We use $\frac{S}{\sqrt{B}}=2.0$ to obtain the exclusion lower bound on the cross section.}%
\label{table:scanning}%
\end{table}

\begin{figure}[t]
\begin{center}
\includegraphics[width=0.9\columnwidth]{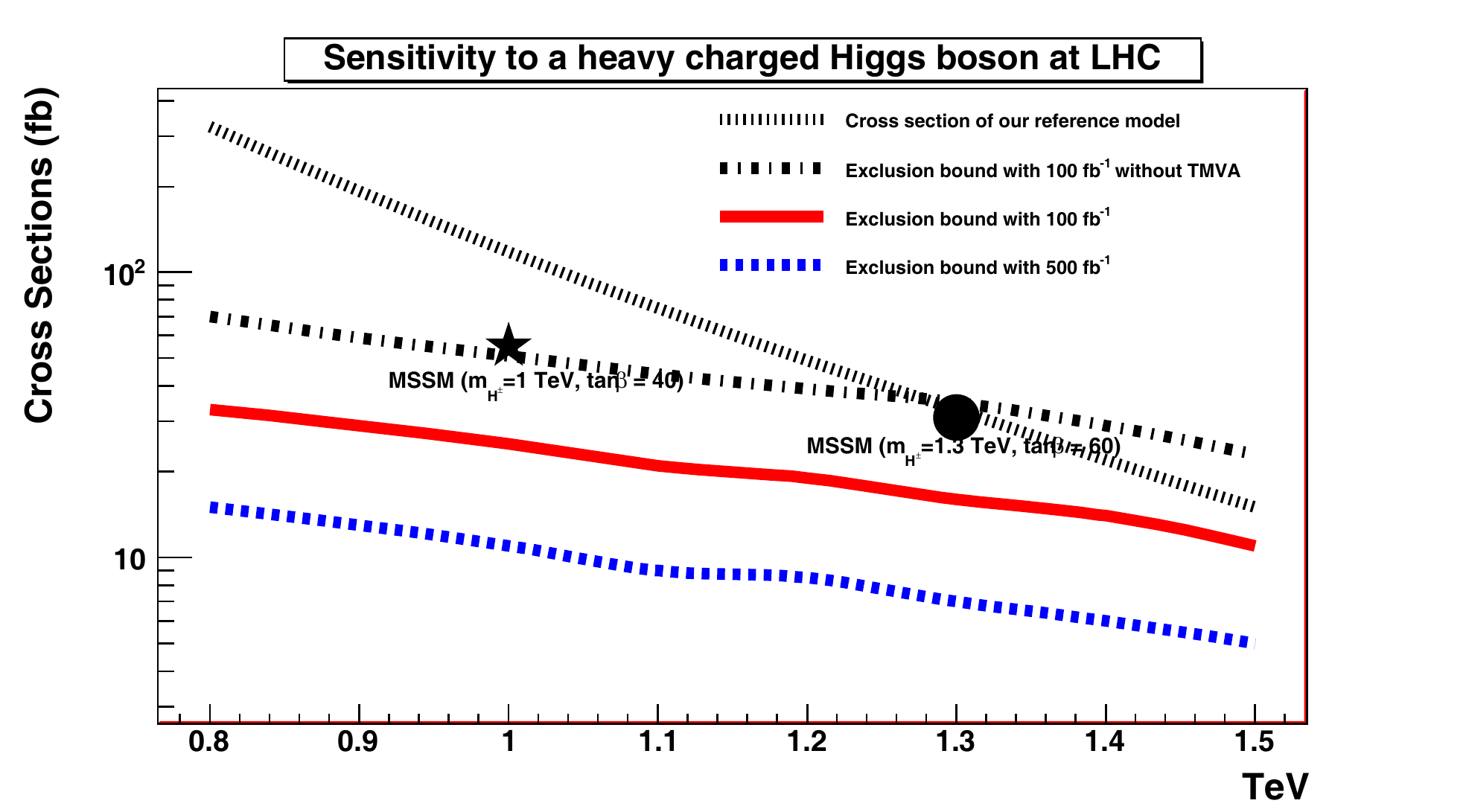}
\caption{The sensitivity curves of LHC to the heavy charged Higgs bosons with two $b$ tagging are shown from both 100 fb$^{-1}$ and 500 fb$^{-1}$ datasets. The solid star and solid circle markers as two example points of MSSM are also shown.}
\label{fig:sen}
\end{center}
\end{figure}

Sensitivity of LHC to heavy charged Higgs boson in the $m_{H^\pm}$-$\sigma(pp \to H^{\pm} t)$ plane are investigated by scanning the charged Higgs boson mass $m_{H^\pm}$ from $800$ GeV to $1.5$ TeV with a step 100 GeV. For each Higgs boson mass, the cone parameter $R$ in the JH-tagger method with the best selection efficiency for signal is taken. The sensitivity reachable at LHC with 14 TeV collision energy with two $b$ taggings is provided in Table \ref{table:scanning} and is displayed in Fig. (\ref{fig:sen}). Roughly speaking, the TMVA can improve the sensitivity by a factor $100\%$. In Fig. (\ref{fig:sen}), we show bounds from both 100 fb$^{-1}$ and 500 fb$^{-1}$ integrated luminosity. Two sample points of MSSM (marked with a solid star and a solid circle, respectively) are also shown.

It is found that the signal and background separation can be improved when the mass of the charged Higgs boson increases. From Fig. \ref{fig:sen}, it is reliable to conclude that the whole mass range of heavy charged Higgs boson in our study can be feasible at the LHC.

\section{Conclusions}
\label{sec:conclusions}
In this paper, we have studied the full hadronic decay mode of the heavy charged Higgs production process $pp \rightarrow t_2H^- \rightarrow t_2\bar{t}_1b_3$. The main task to reconstruct the heavy charged Higgs boson in association with a non-boosted top comes from the large combinatorics and the huge QCD background events. By using the $b$ taggings and top tagger methods to suppress the background from the SM, we propose a hybrid-R reconstruction method to fully reconstruct the signal.

Our hybrid-R reconstruction method exploits the advantages of a larger and a smaller cone size of jet algorithms and can greatly reduce the combinatorics. As demonstrated in our analysis, this method is helpful to confirm the signature as well as to separate the signal from the background.

A set of delicate kinematic observables are applied to the multi-variate analysis. Compared with the simple cut method, an improvement around $100\% $ in the $S/\sqrt{B}$ can be achieved. The sensitivity of the charged Higgs boson mass $m_{H^\pm}$ from $800$ GeV to $1.5$ TeV is also explored and it is found that the LHC can probe the heavy charged Higgs boson mass up to 1.5 TeV. When the integrated luminosity is higher, even heavier mass region of the charged Higgs boson can be reachable.

In this paper, we chose a reference THDM model as an example to perform a concrete study and the cross section of heavy charged Higgs boson is approximately proportional to $Y_L^2+Y_R^2$. To apply our results to other models, like the supersymmetry model, one can rescale the cross section and the significance in term of the model parameters.

In the evaluation of cross section of signal, we have included the process $gb\rightarrow \bar{b}tH^-$ which is a part of QCD next-to-leading (NLO) order correction of process $gb\rightarrow tH^-$and we only assume the K-factor as $1.5$ here. A lot of works on high order QCD corrections and SUSY corrections for process $bg\rightarrow tH^-$ have been done in the past, including QCD NLO calculations
\cite{QCDNLO,SUSYQCDNLO}, SUSY-QCD NLO calculations \cite{SUSYQCDNLO}, Yukawa and SUSY electroweak corrections \cite{EWNLO} as well as high-order soft gluon corrections \cite{NNLO,NNNLO,Kfactor}. These calculations show the reduction of scale dependence and a significant enhancement to lowest order result when high order corrections are considered.
For $\mu=m_{H^\pm}$, $tan \beta=30$ and $m_{H^\pm}=800$GeV (1 TeV), the K-factor in QCD next-to-next-to-next-to leading order soft gluon threshold correction at next-to-leading logarithmic accuracy is 1.75 (1.81) \cite{NNNLO,Kfactor}. Therefore when our results are applied to other models, the significance can be better.

Obviously, our hybrid-R reconstruction method can be generalized to cases where some jets are decay products of one highly boosted massive particle while some of others
come from a massive particle near the threshold. Such cases can exist in some new physics model where the production of a new heavy particles X in association with a SM
massive particle Y, where X decays to boosted massive particles of the SM, such as $W$ bosons,  $Z$ bosons, $H$ bosons or top quarks. For examples,
the hybrid-R reconstruction method can be applied to the processes $pp \rightarrow W^{'} H\rightarrow tbH$ and $pp \rightarrow t^{'} W \rightarrow WbW$
 (where $W^{'}$ and $t^{'}$ represent a new gauge boson and a new up-type quark, respectively).
Additionally, our analysis can be extended to the semi-leptonic and di-leptoinc modes \cite{HClepYY} and to the intermediate heavy charged Higgs boson region (say from 300 GeV to 800 GeV) with the help of HEPToptagger.


\acknowledgments{
We would like to thank C.P. Yuan and S. H. Zhu for helpful discussions. This work was supported in part by the National Science Foundation of China under Grant Nos.  11175251. SY would like to thank the hospitality of the ``2011 IPMU-YITP School and Workshop on Monte Carlo Tools for LHC", and ``2011 CTEQ Workshop and Conference" during this work. The lectures and discussions on these activities are helpful to this work.
}

\end{document}